\documentclass[fleqn,usenatbib]{mnras}

\usepackage{newtxtext,newtxmath}

\usepackage[T1]{fontenc}
\usepackage{ae,aecompl}

\usepackage{graphicx}	
\usepackage{amsmath}	
\usepackage{amssymb}	
\usepackage[xindy]{glossaries}
\glsdisablehyper
\usepackage{subcaption}
\usepackage{tabularx}
\usepackage[dvipsnames]{xcolor}
\usepackage[normalem]{ulem}
\usepackage{siunitx}
\usepackage{appendix}

\newacronym{acf}{ACF}{Auto-Correlation Function}
\newacronym{askap}{ASKAP}{Australian Square-Kilometre Array Telescope Pathfinder}
\newacronym{arsr}{ARSR}{Air Route Surveillance Radar}
\newacronym{bl}{BL}{Breakthrough Listen}
\newacronym{bpsk}{BPSK}{Binary Phase-Shift Keying}
\newacronym{dm}{DM}{Dispersion Measure}
\newacronym{frb}{FRB}{Fast Radio Burst}
\newacronym{fwhm}{FWHM}{Full-Width at Half-Maximum}
\newacronym{geo}{GEO}{Geosynchronous Orbit}
\newacronym{gps}{GPS}{Global Positioning System}
\newacronym{grb}{GRB}{Gamma Ray Burst}
\newacronym{htru}{HTRU}{High-Time Resolution Universe}
\newacronym{igm}{IGM}{Intergalactic Medium}
\newacronym{ixr}{IXR}{Intrinsic Cross-Polarization Ratio}
\newacronym{ism}{ISM}{Interstellar Medium}
\newacronym{leo}{LEO}{Low-Earth Orbit}
\newacronym{meo}{MEO}{Medium Earth Orbit}
\newacronym{pfb}{PFB}{polyphase filterbank}
\newacronym{rfi}{RFI}{Radio-frequency Interference}
\newacronym{rm}{RM}{Rotation Measure}
\newacronym{sefd}{SEFD}{System Equivalent Flux Density}
\newacronym{superb}{SUPERB}{SUrvey for Pulsars and Extragalactic Radio Bursts}
\newacronym{snr}{S/N}{signal-to-noise Ratio}
\newacronym{sps}{SPS}{Single Pulse Search}
\newacronym{tle}{TLE}{Two-Line Element Set}
\newacronym{usaf}{USAF}{United States Air Force}

\newcommand{\frbmjd}{58178.3159487}
\newcommand{\frbtime}{07:34:57.969 }

\title[FRB\,180301]{A Fast Radio Burst with frequency-dependent polarization \\  
detected during Breakthrough Listen observations}

\newcommand{\UCB}{Department of Astronomy,  University of California Berkeley, Berkeley CA 94720}
\newcommand{\SSL}{Space Sciences Laboratory, University of California, Berkeley, Berkeley CA 94720}
\newcommand{\SWIN}{Centre for Astrophysics \& Supercomputing, Swinburne University of Technology, Hawthorn, VIC 3122, Australia}

\newcommand{\OXF}{Astronomy Department, University of Oxford, Keble Rd, Oxford, OX13RH, United Kingdom}
\newcommand{\NIJ}{Department of Astrophysics/IMAPP,Radboud University, Nijmegen, Netherlands}
\newcommand{\ATNF}{Australia Telescope National Facility, CSIRO, PO Box 76, Epping, NSW 1710, Australia}
\newcommand{\ICRAR}{International Centre for Radio Astronomy Research, Curtin University, Bentley, WA 6102, Australia}
\newcommand{\KZA}{University of Malta, Institute of Space Sciences and Astronomy}

\author[D. C. Price et al.]{D. C. Price$^{1,2}$,
G. Foster$^{3,2}$,
M. Geyer$^{4}$,
W. van Straten$^{5}$,
V. Gajjar$^{2,6}$,
G. Hellbourg$^{2,7}$,
\and
A. Karastergiou$^{3,22,23}$,
E. F. Keane$^{8}$,
A. P. V. Siemion$^{2,9,10,11}$,
I. Arcavi$^{12,13,14,15}$,
\and
R. Bhat$^{7}$,
M. Caleb$^{24}$,
S-W. Chang$^{16,17}$,
S. Croft$^{2}$,
D. DeBoer$^{2}$,
I. de Pater$^{2}$,
J. Drew$^{18}$,
\and
J. E. Enriquez$^{2,9}$,
W. Farah$^{1}$,
N. Gizani$^{2,19}$,
J. A. Green$^{20}$,
H. Isaacson$^{2}$,
J. Hickish$^{2}$,
\and
A. Jameson$^{1,17}$,
M. Lebofsky$^{2}$,
D. H. E. MacMahon$^{2}$,
A. M\"oller$^{16,17}$,
C. A. Onken$^{16,17}$,
\and
E. Petroff$^{21}$,
D. Werthimer$^{2,6}$,
C. Wolf$^{16,17}$,
S. P. Worden$^{18}$,
Y. G. Zhang$^{2}$,
\\
$^{1}$\SWIN\\
$^{2}$\UCB\\
$^{3}$\OXF\\
$^{4}$SARAO, 2 Fir Street, Black River Park, Observatory, Cape Town, 7925, South Africa\\
$^{5}$Institute for Radio Astronomy and Space Research, Auckland University of Technology, PB 92006, Auckland 1142, New Zealand\\
$^{6}$\SSL\\
$^{7}$\ICRAR\\
$^{8}$SKA Organisation, Jodrell Bank Observatory, Macclesfield, Cheshire, SK11 9DL, UK\\
$^{9}$\NIJ\\
$^{10}$SETI Institute, Mountain View, California, USA\\
$^{11}$\KZA\\
$^{12}$Department of Physics, University of California, Santa Barbara, CA 93106-9530, USA\\
$^{13}$Las Cumbres Observatory, 6740 Cortona Dr Ste 102, Goleta, CA 93117-5575, USA\\
$^{14}$Einstein Fellow\\
$^{15}$The Raymond and Beverly Sackler School of Physics and Astronomy, Tel Aviv University, Tel Aviv 69978, Israel\\
$^{16}$Research School of Astronomy and Astrophysics, The Australian National University, Canberra, ACT 2611, Australia\\
$^{17}$ARC Centre of Excellence for All-sky Astrophysics (CAASTRO), School of Physics, The University of Sydney, Sydney, NSW 2006, Australia\\
$^{18}$The Breakthrough Initiatives, NASA Research Park, Bld. 18, Moffett Field, CA, 94035, USA\\
$^{19}$School of Science and Technology, Hellenic Open University, Patra, Greece\\
$^{20}$\ATNF\\
$^{21}$Anton Pannekoek Institute for Astronomy, University of Amsterdam, Science Park 904, 1098 XH Amsterdam, The Netherlands\\
$^{22}$Physics Department, University of the Western Cape, Cape Town 7535, South Africa\\ 
$^{23}$Department of Physics and Electronics, Rhodes University, PO Box 94, Grahamstown 6140, South Africa\\
$^{24}$Jodrell Bank Centre for Astrophysics, the University of Manchester, Manchester, UK
}

\date{Accepted XXX. Received YYY; in original form ZZZ}

\pubyear{2018}

\begin{document}
\label{firstpage}
\pagerange{\pageref{firstpage}--\pageref{lastpage}}
\maketitle

\begin{abstract}
Here, we report on the detection and verification of Fast Radio Burst
FRB\,180301, which occurred on UTC 2018 March 1 during the Breakthrough Listen
observations with the Parkes telescope. Full-polarization voltage data of the
detection were captured---a first for non-repeating FRBs---allowing for coherent
de-dispersion and additional verification tests. The coherently de-dispersed
dynamic spectrum of FRB\,180301 shows complex, polarized frequency structure
over a small fractional bandwidth. As FRB 180301 was detected close to the
geosynchronous satellite band during a time of known 1--2\,GHz satellite
transmissions, we consider whether the burst was due to radio interference
emitted or reflected from an orbiting object.  Based on the preponderance of our
verification tests, we find that FRB\,180301 is likely of astrophysical origin,
but caution that anthropogenic sources cannot conclusively be ruled out.  
\end{abstract}

\begin{keywords}
radio continuum: transients -- methods: data analysis -- methods: observational
\end{keywords}



\section{Introduction}

\glspl{frb}, first reported by \citet{Lorimer:2007}, are a now-routinely
detected---but nonetheless rare---class of transient radio sources of inferred
extragalactic origin \citep[e.g.][]{Thornton2013,2017MNRAS.468.3746C,Ravi:2019};
see FRBCAT\footnote{\url{http://frbcat.org/}} for an up-to-date catalogue
\citep{2016PASA...33...45P}. Identifying the sources of \glspl{frb} and
understanding their emission mechanisms is an area of active research within
astronomy. Given their extreme luminosities (isotropic burst energies
$>10^{40}$\,erg, \citealt{Dolag:2015}), and their inferred cosmological
distances, \glspl{frb} could be used as cosmological probes
\citep{Zhou2014PhRvD, Deng2014ApJ, Walters2018ApJ, Keane:2018NA}. 

To unambiguously prove an extragalactic origin of an \gls{frb}, many surveys are
focused on using interferometric arrays to localize the source to host galaxies
at the time of detection
\citep[e.g.][]{Law:2015realfast,Spolaor:2016vfastr,2017ApJ...841L..12B,2017MNRAS.468.3746C}.
Further strides towards understanding the nature of \glspl{frb} come from more
complete sampling of frequency space
\citep{2017ApJ...844..140C,2018arXiv180311235T,Amiri:2019}, and capturing
high-time resolution voltage data of a detection \citep{2018MNRAS.478.1209F}. 

To date, only two \glspl{frb} have been shown to repeat: FRB\,121102
\citep{Spitler:2016}, and FRB\,180814.J0422+73 \citep{Amiri:2019r}. The
repetition of FRB\,121102 allowed interferometric localization and host galaxy
identification via follow-up observations \citep{Chatterjee:2017}. These
observations unambiguously showed FRB\,121102 to be astrophysical in origin, at
a distance $z\lesssim 0.192$.  While FRB\,121102 appears to have active and
non-active phases, no underlying periodicity has been detected
\citep{Zhang:2018ml}. Efforts to localize the recently-discovered
FRB\,180814.J0422+73 are ongoing \citep{Amiri:2019r}.

Nevertheless, it cannot yet completely be ruled out that some fraction of
\glspl{frb} are false-positives from \gls{rfi} as FRB-like \gls{rfi} is known to
exist. A subset of FRB-like signals, dubbed `perytons', showed signs of
near-field terrestrial origin \citep{2011ApJ...727...18B}; eventually, these
signals were shown to be caused by an on-site interferer \citep{Petroff:2015}.
A variety of of FRB-like \gls{rfi} are presented in
\citet{2018MNRAS.481.2612F}, along with a verification framework aimed at
limiting false positives.  (In practice, all one can do is perform as many
verification tests as the data allow.)  The FRBs reported so far pass all of the
tests that it has been possible to perform.  However, as we collect increasingly
rich information, we can be increasignly rigorous in our verification. This is
important as the understanding of the diverse manifestations of RFI is
incomplete.

\glspl{frb} display varying frequency and polarization characteristics that may
be intrinsic or extrinsic to the emission mechanism. Several FRB
events---including FRB\,110523 \citep{2015Natur.528..523M}, FRB\,170827
\citep{2018MNRAS.478.1209F}, and FRB\,121102 \citep{2018Natur.553..182M,
gsp+18}---  show spectral modulation on scales of order $\sim$1 MHz.
\citet{Shannon:2018} recently reported 20 \glspl{frb} detected with the
\gls{askap}, all of which exhibit spectral modulation. Similarly, spectral
modulation is also apparent in the 13 bursts detected using the Canadian
Hydrogen Intensity Mapping Experiment \citep[CHIME,][]{Amiri:2019}.  If
intrinsic to the source, the frequency-modulated emission is distinctly
different from the broadband emission associated with possible progenitors such
as (young) pulsars and magnetars \citep{Jankowski:2018}. In most cases, however,
the modulated emission has been attributed to propagation effects, namely inter-
and intra-galactic scintillation.

Polarization properties and Faraday \gls{rm} have also been measured for a
number of \glspl{frb} \citep{Caleb:2018}. Large polarization fractions imply the
existence of strong magnetic fields in the progenitor or its immediate
environment, while large \glspl{rm} imply strong magnetic fields along the line
of sight.  A significant linear polarization fraction was reported for
FRBs\,110523, 150215, 150418, 150807, 151230 and 160102; circular polarization,
while less common, is exhibited by FRB\,140514, 150215, and 160102; see Table 1
of \citep{Caleb:2018} for a summary of polarization properties.  Measurements of
\glspl{rm}  (inconsistent with zero) are reported for FRBs\,110523, 150807, and
160102.  FRB\,121102 remarkably exhibits an \gls{rm} in excess of
$10^5$\,rad\,m$^{-2}$ and appears to evolve with time
\citep{2018Natur.553..182M, gsp+18}.

As the number of detected \glspl{frb} increases, it may become apparent that
there are distinct classes, and that the broadly-varying burst characteristics
are due to different emission mechanisms. Statistically robust relationships
between observed quantities may also become apparent.  For example, from
analysis events detected with Parkes and of 20 \gls{frb} events detected with
\gls{askap}, \citet{Shannon:2018} report a relationship between dispersion and
brightness.  A relationship between dispersion and scattering also appears to
hold \citep[e.g.][]{Amiri:2019}.

Here, we report the detection of a highly-polarized \gls{frb}, henceforth
FRB\,180301. The \gls{frb} was detected during Breakthrough Listen
bservations of the Galactic plane \citep{Worden:2017aa,2017PASP..129e4501I}.
The remainder of this paper is structured as follows. The
detection of FRB\,180301 is described in \S \,\ref{sec:detect}, and its detailed
verification in \S  \ref{sec:verify}, following procedures set out in
\cite{2018MNRAS.481.2612F}. In particular, we consider evidence for the event
being related to a geosynchronous satellite. Details of follow-up observations
are given in \S\,\ref{sec:follow-up}. In \S\,\ref{sec:discuss} we discuss our
findings, with conclusions drawn in \S\,\ref{sec:conclude}.

\section{Observations}
\label{sec:detect}

FRB\,180301 was detected on UTC 2018 March 1 at \frbtime (MJD \frbmjd,
referenced at 1415\,MHz) during \gls{bl} observations with 
the CSIRO Parkes 64-m radio telescope (see
Appendix\,\ref{BL} for program details).  The event occurred in beam 03 of the
21-cm multibeam receiver, the J2000 coordinates of the beam centre during the
event were (06:12:43.4, 04:33:44.8), with corresponding Galactic $(l, b)$
coordinates ($204.412^{\circ}$, $-6.481^{\circ}$); the full-width half maximum
width of the beam is $\sim$14.1 arcmin. After initial verification, an
\emph{Astronomer's Telegram} was issued to allow for immediate follow-up by
other facilities \citep{atel11376}. 

FRB\,180301 was detected in real-time using the Berkeley-Parkes-Swinburne
Recorder (BPSR) and HI-Pulsar (HIPSR) system \citep{Keith:2010htru,
Price:2016jai, Keane2015}, which is configured to run in parallel with the
\gls{bl} digital recorder \citep{2018PASA...35...41P}. Running BPSR is an
addition to the original mode of operation, to allow commensal science during BL
observations.  BPSR records dynamic spectra with a time resolution of
\SI{64}{\micro\second}, and  channel resolution of \SI{390.625}{\kilo\hertz},
spanning the receiver's usable band 1.182--1.522\,GHz.  The BPSR system performs
a brute-force incoherent de-dispersion search \citep{Barsdell:2012}  in
real-time, alerting observers via email to candidate FRB events. The BPSR and BL
systems are fully independent, and were both recording data at the time of the
event. 

The BPSR incoherent search pipeline identified FRB\,180301 as a candidate FRB
event with a \gls{snr} of $\sim$16 at a \gls{dm} of 520~pc~cm$^{-3}$, using a 
2.048 ms boxcar filter; a dynamic spectrum plot of this detection is shown in Figure
10 of \cite{2018PASA...35...41P}. Shortly after visual inspection of the
candidate signal, we interrupted regular \gls{bl} observations, and undertook
follow-up observations and calibration procedures.

FRB\,180301 was detected in data from both BPSR and the \gls{bl} recorder; for
the analysis presented in this paper, we primarily use \gls{bl} data products.
The BL data recorder system records complex voltage products over 308\,MHz of
bandwidth (1.2075--1.5155\,GHz) to disk in GUPPI raw format
\citep{Ford:2010p8485}, for each of the multibeam receiver's 13 beams. The
voltages are coarsely channelized into sub-bands of width 3.5\,MHz using a
critically-sampled \gls{pfb}; further instrument details may be found in
\citet{2018PASA...35...41P}. The Parkes \gls{bl} data recorder shares system
design with the \gls{bl} data recorder at the Green Bank telescope
\citep{2018PASP..130d4502M}, used in the detection of  FRB events from
FRB\,121102 over 4 to 8\,GHz \citep{gsp+18,Zhang:2018ml}.

\begin{figure*}
    \includegraphics[width=0.85\linewidth]{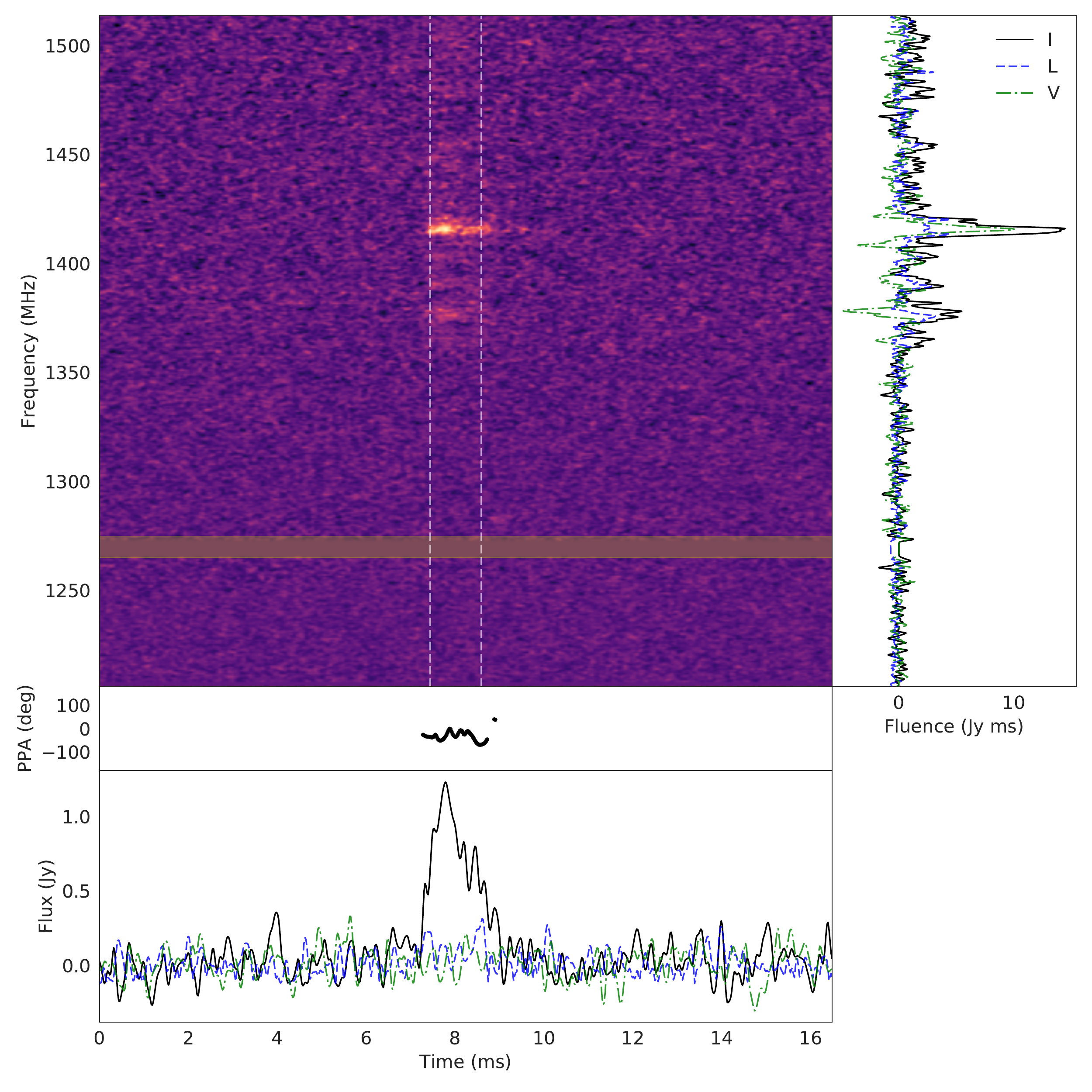}
    \caption{FRB\,180301 coherently de-dispersed with a DM of $522\pm
    5$\,pc\,cm$^{-3}$ to a resolution of \SI{22}{\micro\second} and
    $109.375$~kHz, with a Gaussian noise filter (\SI{88}{\micro\second},
    220\,kHz) applied. No rotation measure correction has been applied.
    Central figure is the Stokes I dynamic spectrum.  Summed spectrum between
    the white dashed lines is shown on the right; Stokes\,L and Stokes\,V are
    plotted in blue dashed and green dashed lines, respectively.  The summed
    profile and polarization position angle are plotted in the bottom of the
    figure. Persistent \gls{rfi} around $\sim$1270\,MHz has been flagged.
    }
    \label{fig:dynamic_spectrum}
\end{figure*}

Nyquist-sampled dual-polarization voltage-level products were recorded for all
receiver beams during the detection of FRB\,180301.  These data can be
coherently de-dispersed to remove temporal smearing, increasing the \gls{snr} to
20, and allow finer control of the time and frequency resolution of derived
dynamic spectrum. A coherently de-dispersed dynamic spectrum, with a time and
frequency resolution of \SI{22}{\micro\second} and $109.375$\,kHz, is shown in
Figure \ref{fig:dynamic_spectrum}; a summary of the FRB\,180301 detection and
its derived properties is given in Table\,\ref{tbl:properties}.

\section{Analysis}
\label{sec:verify}

\begin{figure*}
    \includegraphics[width=0.8\linewidth]{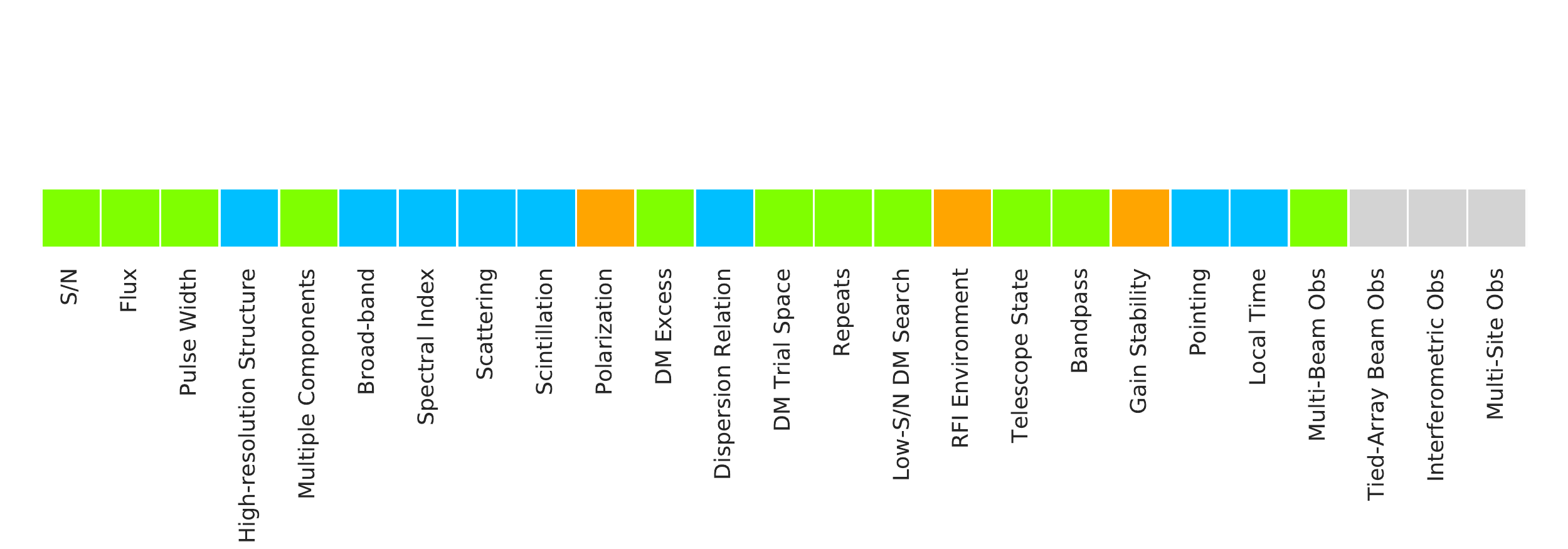}
    \caption{Results heat map of verification tests \citep{2018MNRAS.481.2612F}
    for FRB\,180301. Green indicates the test result is identical to the
    prototypical FRB or an ideal observation. Blue indicates the test result is
    similar to the ideal test result, but not identical. Orange indicates the
    test result is significantly different from the ideal result, and could
    indicate the FRB is terrestrial. Gray indicates a test which was not valid
    for the observation.
    }
    \label{fig:heatmap}
\end{figure*}

Here, we detail observed burst characteristics, and apply the tests presented in
\cite{2018MNRAS.481.2612F} as a framework to verify FRB\,180301 as
astrophysical. A heat map of these test results is shown in Figure
\ref{fig:heatmap}, the individual tests are discussed throughout this section.
Flux calibration was performed by observing calibrator PKS1934$-$638 in each
beam at the beginning of the observation, and polarization calibration was
performed with noise diode reference observations. Beam 03 had a
frequency-averaged \gls{sefd} of $\sim$37\,Jy. This is higher than the central
beam which typically has an \gls{sefd} of $\sim$30\,Jy, predominantly due to
dish optics \citep{StaveleySmith:1996mb}.  Assuming the pulse occurred near the
centre of beam 03, the frequency-averaged profile has a peak flux of $1.23$\,Jy.
This is more than twice the originally reported peak flux since using coherent
de-dispersion has reduced smearing and allowed for the main pulse to be time
resolved.

\begin{table}
    \centering
    \begin{tabular}{ll}
    Property & Value \\ 
    \hline\hline
    Identifier & FRB 180301 \\
    UTC Date & 2018-03-01 \\
    UTC Time & \frbtime \\
    Local Time (AEDT) & 18:34:57.969 \\
    Modified Julian Date & \frbmjd \\
    \hline
    Telescope / Receiver & Parkes 21-cm multibeam \\
    Observing Band & $1.2075$--$1.5155$ GHz \\
    Local Coords (alt, az) & $41.4^{\circ}$, $45.6^{\circ}$\\
    Celestial (J2000) ($\alpha, \delta$) &  $06^{\rm{h}}12^{\rm{m}}43.4{\rm{s}}$, $04^{\rm{d}}33^{\rm{m}}45.4^{\rm{s}}$ \\
    Galactic ($l$, $b$) &  $204.412^{\circ}$, $-6.481^{\circ}$ \\
    \hline
    Detection \gls{snr} & 16 \\
    Optimal \gls{snr} & 20 \\
    Peak flux density (Jy) & 1.2 $\pm \, 0.1$\\
    DM (pc~cm$^{-3}$) & $522 \pm 5$ \\
    DM index & $-1.9 \pm 0.1$ \\
    Pulse width (W10) (ms)$^\dagger$ & $2.18 \pm 0.06$ \\
    Pulse width (W10) (ms)$^\ddagger$ & $0.74 \pm 0.05$ \\
    $\tau_{\textrm{scattering}}$ (ms)$^\ddagger$ & $0.71 \pm 0.03$ \\
    RM (rad m$^{-2}$)$^\star$ & $-3163 \pm 20$ \\ 
    \hline
    \end{tabular}
    \caption{Summary of FRB\,180301 detection and derived properties. $\dagger$
    Pulse width fit for a Gaussian component model. $\ddagger$ Pulse width and
    scattering timescale for a scattered Gaussian component model. $\star$
    Rotation measure assuming polarization characteristics are due to Faraday
    rotation; see Sec.~\ref{sec:verify} for details.}
    \label{tbl:properties}
\end{table}

\subsection{Radio-Frequency Interference}

\begin{figure*}
    \centering
    \begin{subfigure}[b]{0.45\textwidth}
        \centering
        \includegraphics[width=1.0\linewidth]{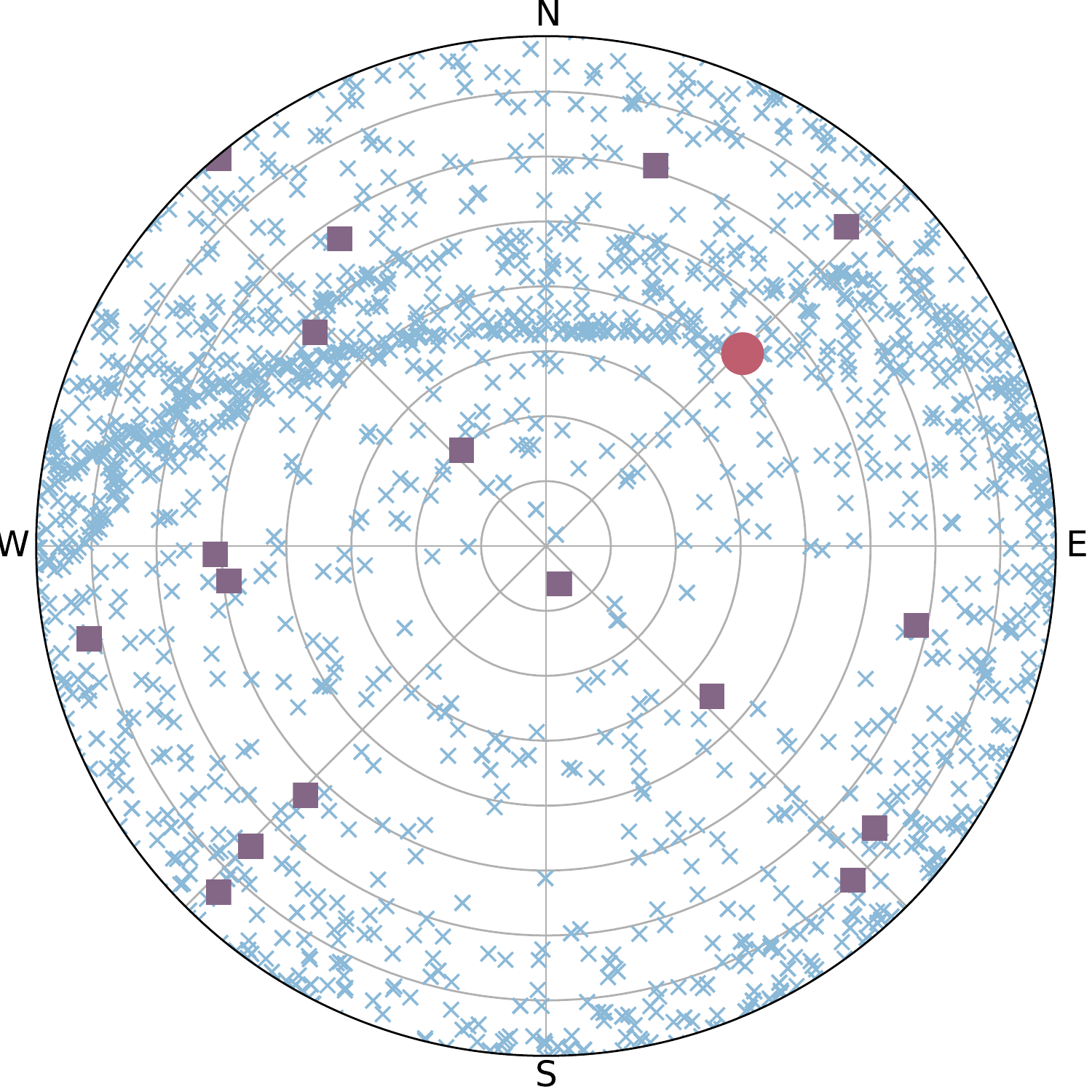}
        \caption{}
        \label{fig:sat_pos}
    \end{subfigure}%
    ~ 
    \begin{subfigure}[b]{0.45\textwidth}
        \centering
        \includegraphics[width=1.0\linewidth]{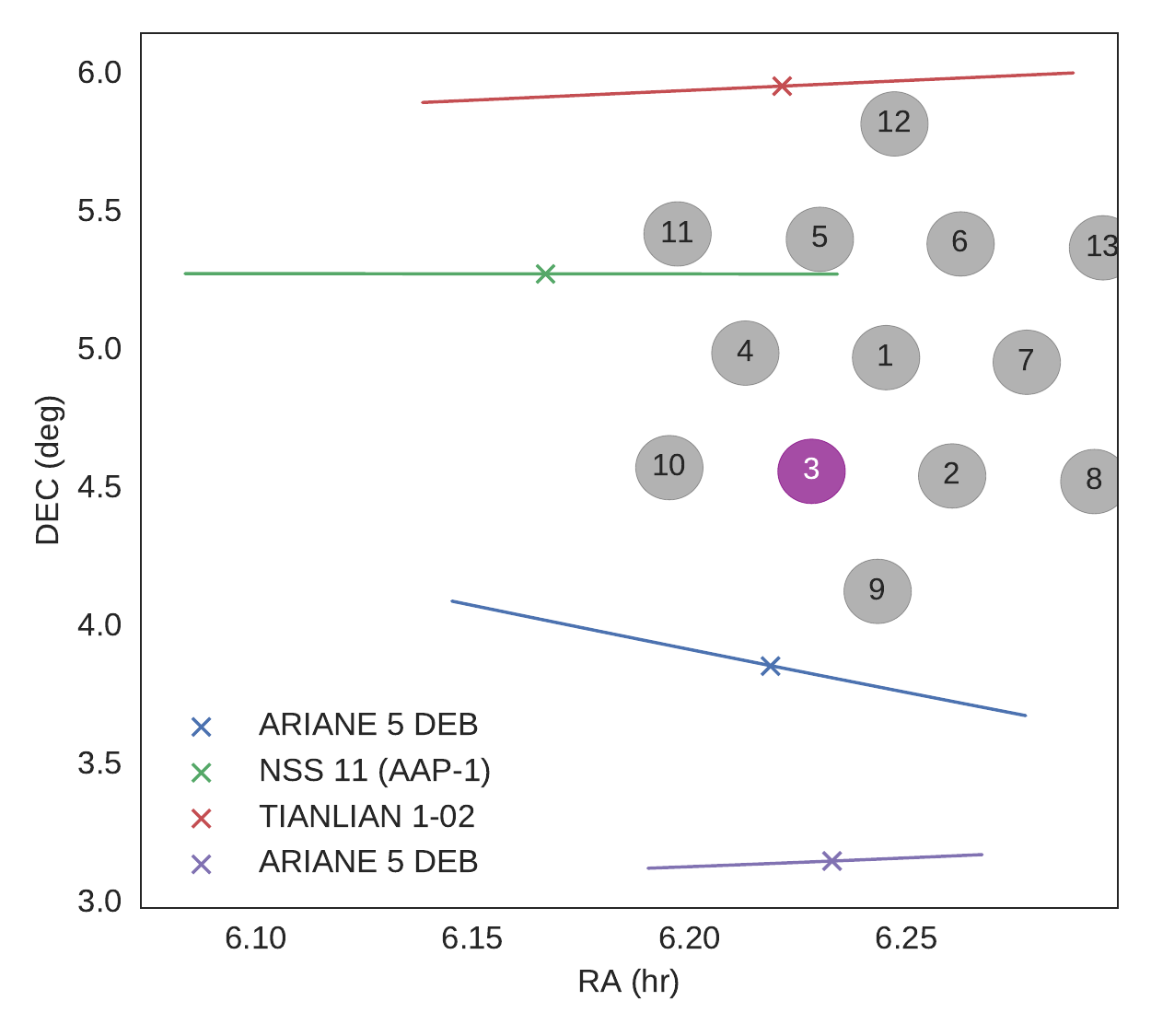}
        \caption{}
        \label{fig:sat_paths}
    \end{subfigure}
    \caption{(a) Location of all publicly-listed satellites above the horizon
    for Parkes at UTC 2018-03-01 07:34:19. FRB\,180301 is shown as a red circle;
    GPS satellites are shown as purple squares. (b) Satellites (NSS-11, TIANLIAN
    1-02) and Ariane 5 rocket debris paths near the pointing of the multibeam
    receivers (circles) within a $\pm 5$~minute window of the \gls{frb}
    detection in beam 3 (purple). Crosses mark the position of the satellites at
    time of detection.
    }
\end{figure*}

To assure the quality of the detection, we investigated the state of the
telescope and \gls{rfi} environment. At the beginning of the observation,
calibrator source PKS1934-638 was observed in each beam at the expected
\gls{snr}. The local time during detection was early evening, meaning that the
visitor's centre is closed, and visitor-related \gls{rfi} sources are fewer. The
overall \gls{rfi} was low during the time of the detection; the Parkes RFI
monitor\footnote{\url{https://www.narrabri.atnf.csiro.au/observing/rfi/monitor/rfi_monitor.html\#parkes}},
which operates over 0.4--3.0\,GHz, does not show any notable RFI events during the observation period.
Persistent \gls{rfi} associated with Global Navigation Satellite System (GNSS)
transmitters around 1207\,MHz, 1246\,MHz, and 1270\,MHz regularly seen in Parkes
data were present, and removed during calibration. As flux of the pulse was not
seen in the lower half of the band, there is no expectation that this \gls{rfi}
is the progenitor of the observed pulse. Following the suggestions in \citet{2018MNRAS.481.2612F}, a \gls{dm}-trial search from $-2000$ to
$2000$~pc~cm$^{-3}$ ($\Delta$DM step size of 10~pc~cm$^{-3}$) during the time of
detection revealed no significant \gls{rfi} events.

As the telescope was pointed in the region of the sky where geosynchronous
satellites operate (declinations $\pm 15^{\circ}$)
\citep{anderson15operational}, we extended the low-altitude pointing test in
\cite{2018MNRAS.481.2612F} to also check for the presence of satellites near the
beam (within a few degrees).  Though commercial satellites are not known to
transmit at the frequency of the detected pulse, we still investigated if the
source of the event could be due to a satellite.  Using public \gls{tle} orbital
parameters of tracked satellites we found geosynchronous satellites (NSS-11,
TIANLIAN 1-02) and debris from an Ariane 5 
rocket---a European satellite launch vehicle--near the beams
at the time of the detection (Figure \ref{fig:sat_paths}). NSS-11 is a Ku-band broadcast
satellite\footnote{\url{https://www.ses.com/our-coverage/satellites/355}}
(12--18\,GHz), and Tianlian 1-02 is a Chinese data relay
satellite\footnote{\url{https://nssdc.gsfc.nasa.gov/nmc/spacecraftOrbit.do?id=2011-032A}};
neither is known to transmit in the 1 to 2\,GHz band.  It is possible the event
is related to a government satellite ---for example military satellites are known
to use L-band frequencies--- but a complete record of such satellites is not
made available publicly.

\begin{figure}
    \includegraphics[width=1.0\linewidth]{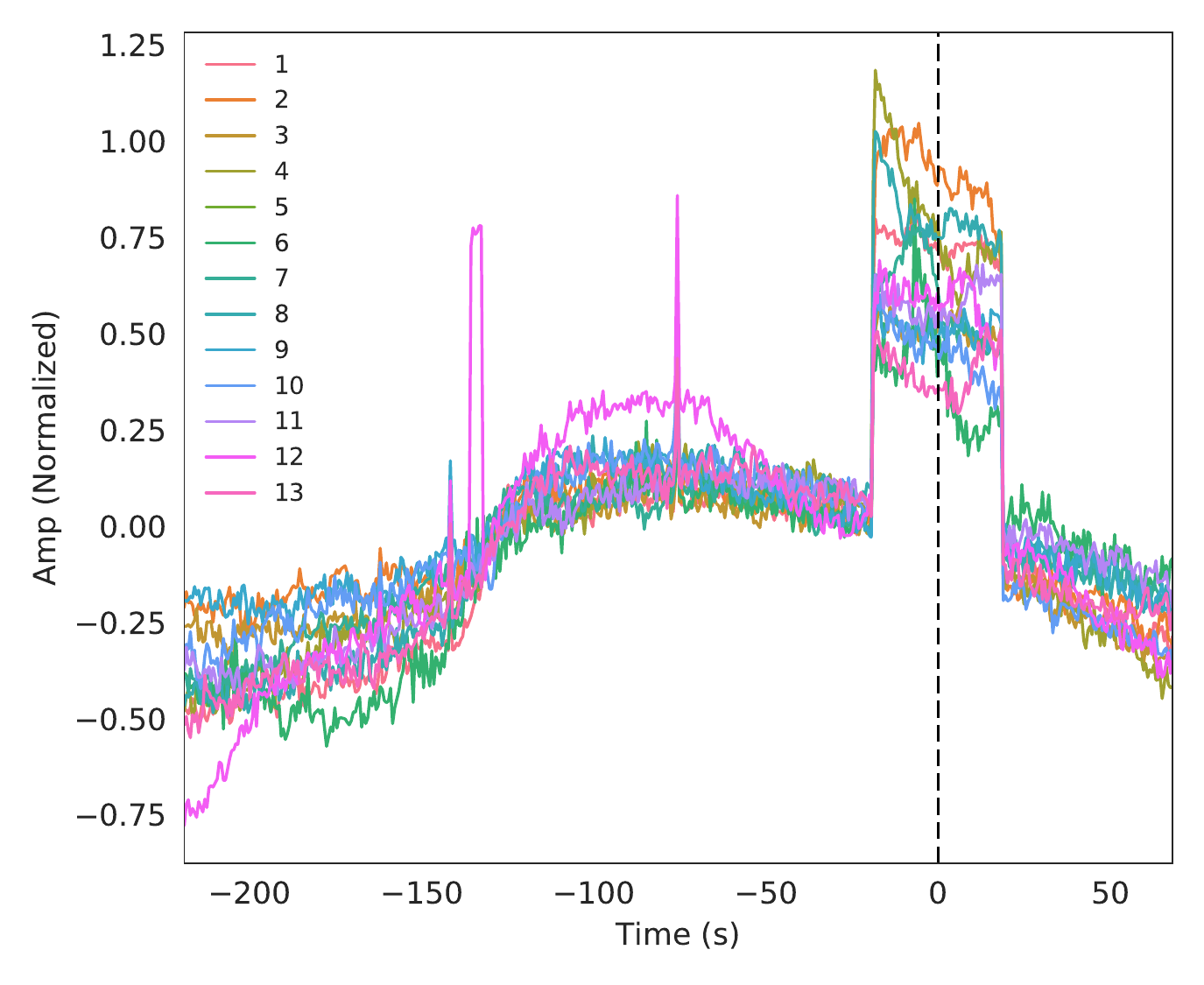}
    \caption{Time series (median removed, normalized) of the total power over
    the extent of the pulse bandwidth (1350 -- 1460~MHz) in each of the 13
    beams. The detection of FRB\,180301 (dashed, black) occurred during a period
    of a GPS L3 test signal transmission.}
    \label{fig:med_res_ts}
\end{figure}

One of the bright regions of the FRB\,180301 dynamic spectrum overlaps with the
\gls{gps} L3 band ($1381.05 \pm 2.5$~MHz), operated by the \gls{usaf}. The L3
band is used as part of the nuclear detection system payload present on every
\gls{gps} satellite. The detection systems are tested quarter-yearly by the
\gls{usaf}\footnote{\url{https://science.nrao.edu/facilities/gbt/observer-alerts/gps-l3-1381-05-2-5-mhz-test-times}}.
During the detection of FRB\,180301 such tests were occurring throughout the
satellite constellation (antenna monitor W8 at the Karl G. Jansky Very Large
Array (VLA) detected use of this band throughout the
day\footnote{\url{http://www.vla.nrao.edu/cgi-bin/rfi.cgi}}).  The detection of
FRB\,180301 occurred at approximately the mid-point of a 30-second period
associated with \gls{gps} L3 signal transmission (Figure \ref{fig:med_res_ts}).
While the total power in Figure \ref{fig:med_res_ts} increases during transmission, 
the L3 signal remains within its specified band, and is too faint 
to appear in the Parkes RFI monitor data.
\gls{gps} satellites are in \gls{meo} and use high-power, wide-beam
transmitters. Though a \gls{gps} satellite was not near the beams (Figure
\ref{fig:sat_paths}), a satellite tens of degrees off from the pointing centre
is still sufficiently powerful to be detected in the side-lobes of all the
beams. We note that this satellite emission could possibly go undetected if the
spectra had been normalized and re-quantized or if only a short period of time
around the burst was examined.  We are not able to determine if this L3
transmission is related to the detected burst, or merely coincident in time.

We attempted to further localize the \gls{frb} by cross-correlating the complex
voltages from each beam with beam 03 to check for beam side-lobe detections, but
no detections were made. Additionally, a low-\gls{snr} search of the others
beams resulted in a non-detection. The brightest spectral structure has a
\gls{snr} of $\sim40$, non-detection of this emission in the neighboring beams
indicates the burst occurred near to the centre of beam 03, or that the
intrinsic flux was very large but occurred in an advantageous far side-lobe of
the beam \citep{2018MNRAS.474.1900M}.  Non-detection in adjacent beams indicates
the source is likely in the far-field of the dish ($\gtrsim 39$~km).  Detection
occurred when the telescope was locally positioned at azimuth and altitude
($45.6^{\circ}, 41.4^{\circ}$). This mid-altitude pointing is far from the
horizon that the pulse is not likely associated with a fixed-position \gls{rfi}
source in the far-field.

To summarize, the \gls{gps} L3 emission was detected in all beams with similar
amplitude (Figure\,\ref{fig:med_res_ts}), presumably through far-beam side-lobes,
whereas the \gls{frb} was detected only in beam 03, likely near the beam centre. This
supports the model that the burst is not directly related to the \gls{gps} L3
emission.

\subsection{Dispersion Measure}

The peak \gls{snr} dispersion measure was fit by performing a coherent
de-dispersion at the original detected dispersion measure of 520~pc~cm$^{-3}$,
then incoherently de-dispersing over a range of $\pm 20$~pc~cm$^{-3}$ in $0.1$
increments. A 2-D Gaussian was fit to this trial \gls{dm} vs. time space to find a peak
at $522 \pm 5$~pc~cm$^{-3}$. The voltage data was then coherently de-dispersed at
this dispersion measure.

A dispersion relation model ($\nu^{\beta}$) was fit to the dispersed pulse
resulting in a best fit relation of $\beta=-1.9 \pm 0.1$.  The error in the
\gls{dm} and the dispersion relation fit range are larger than other \glspl{frb}
detected at Parkes, due to the band-limited nature of the pulse. 
As a point of compariosn, FRBs 110220 and 110703 follow $\beta=-2.003(6)$ and
$\beta=-2.000(6)$, respectively \citep{Thornton2013};
FRB\,140504 follows $\beta=-2.000(4)$  \citep{Petroff:2015pol}.

The line-of-sight Galactic \gls{dm} contribution is 150~pc~cm$^{-3}$ using the
NE2001 model \citep{2002astro.ph..7156C} and 252~pc~cm$^{-3}$ using the YWM16
model \citep{2017ApJ...835...29Y}. The average of these DM model values results
in an excess dispersion of $\sim320$~pc~cm$^{-3}$. An upper limit on the
distance to the host can be determined by assuming the excess dispersion is due
to only the \gls{igm}. Using the \cite{2004MNRAS.348..999I} model results in a
distance $z \lesssim 0.35$ assuming an average line of sight through the
\gls{igm}.  However, there is a large uncertainty in the \gls{igm} dispersion
measure contribution.  FRB\,121102, which appears to be in a dense plasma
environment \citep{2018Natur.553..182M}, has an estimated host \gls{dm}
contribution of $70 - 270$~pc~cm$^{-3}$ and a corresponding distance of $z\sim 0.192$
\citep{2017ApJ...834L...8M}. Assuming the same distance to the host galaxy of
FRB\,180301 would imply a host \gls{dm} contribution of $\sim140$~pc~cm$^{-3}$.

\subsection{Pulse Profile and Scattering Models}

Modelling the pulse profile (Figure \ref{fig:dynamic_spectrum}, lower plot) as a
Gaussian results in a fit width (at 10\% of the peak, W10) of $2.18 \pm
0.06$\,ms.  The fluence over the W10 width is $1.3$\,Jy~ms.  This simple model results in a poor fit to the profile, indicating a more complex model, such as a multi-component or scattered profile model, may be necessary.  We find a best-fit two-component Gaussian with widths $2.41 \pm 0.10$\,ms and $0.88 \pm 0.09$\,ms separated by $0.41$\,ms. Alternatively fitting the (frequency averaged) profile as a Gaussian scattered by an isotropic screen model \citep{2017MNRAS.470.2659G} results in a significantly narrower pulse width
(W10) of $0.74 \pm 0.05$\,ms with a scattering timescale $\tau_{\textrm{scatter}}$ of $0.71 \pm 0.03$\,ms at 1.4~GHz. Comparing the residuals of these two models to normal distributions, by means of a Kolmogorov-Smirnov (KS) test, we conclude that the lower-order scattering model performs best (KS-test p-values of 0.89 and 0.46 respectively).

We performed the same model fit for three regions (of comparable S/N) within
the spectrum: 1370--1410\,MHz, across the bright spectral structure around 1415\,MHz and at a higher frequency interval of 1420--1480\,MHz (see Table
\ref{tbl:scattering}). We find no evidence of a frequency-dependent scattering
timescale.  As there is no clearly preferred pulse profile model, it is possible that
FRB\,180301 is intrinsically asymmetric.


\begin{table}
    \centering
    \begin{tabular}{c|c|c}
    $\nu$ (MHz)    & W10 ($\mu$s)  & $\tau_{\textrm{scatter}}$ ($\mu$s) \\\hline\hline
    1370 -- 1480   & $740 \pm 50$  & $710 \pm 30$    \\\hline
    1370 -- 1410   & $770 \pm 90$  & $650 \pm 50$    \\
    1410 -- 1420   & $710 \pm 50$  & $800 \pm 30$    \\
    1420 -- 1480   & $790 \pm 130$ & $530 \pm 70$   
    \end{tabular}
    \caption{Pulse profile fit for a Gaussian (W10) scattered by an
    isotropic scattering screen ($\tau$) for the extent of the pulse (1370 --
    1480~MHz) and three sub regions of the spectrum.}
    \label{tbl:scattering}
\end{table}

\subsection{Spectro-temporal Structure}

The pulse spectrum (Figure \ref{fig:dynamic_spectrum}, right plot) is band- limited
with narrow frequency features, potentially due to scintillation. The
primary feature of the spectrum is centered around 1415~MHz, this $\sim$10\,MHz
wide feature accounts for a third of the total flux.  The spectrum shows lower
intensity spectral structures between 1350\,MHz and the upper edge of the band
at 1500~MHz. There is no apparent flux below 1350\,MHz. Due to the complex
structure of the spectrum no spectral index model was fit.

\begin{figure}
    \includegraphics[width=1.0\linewidth]{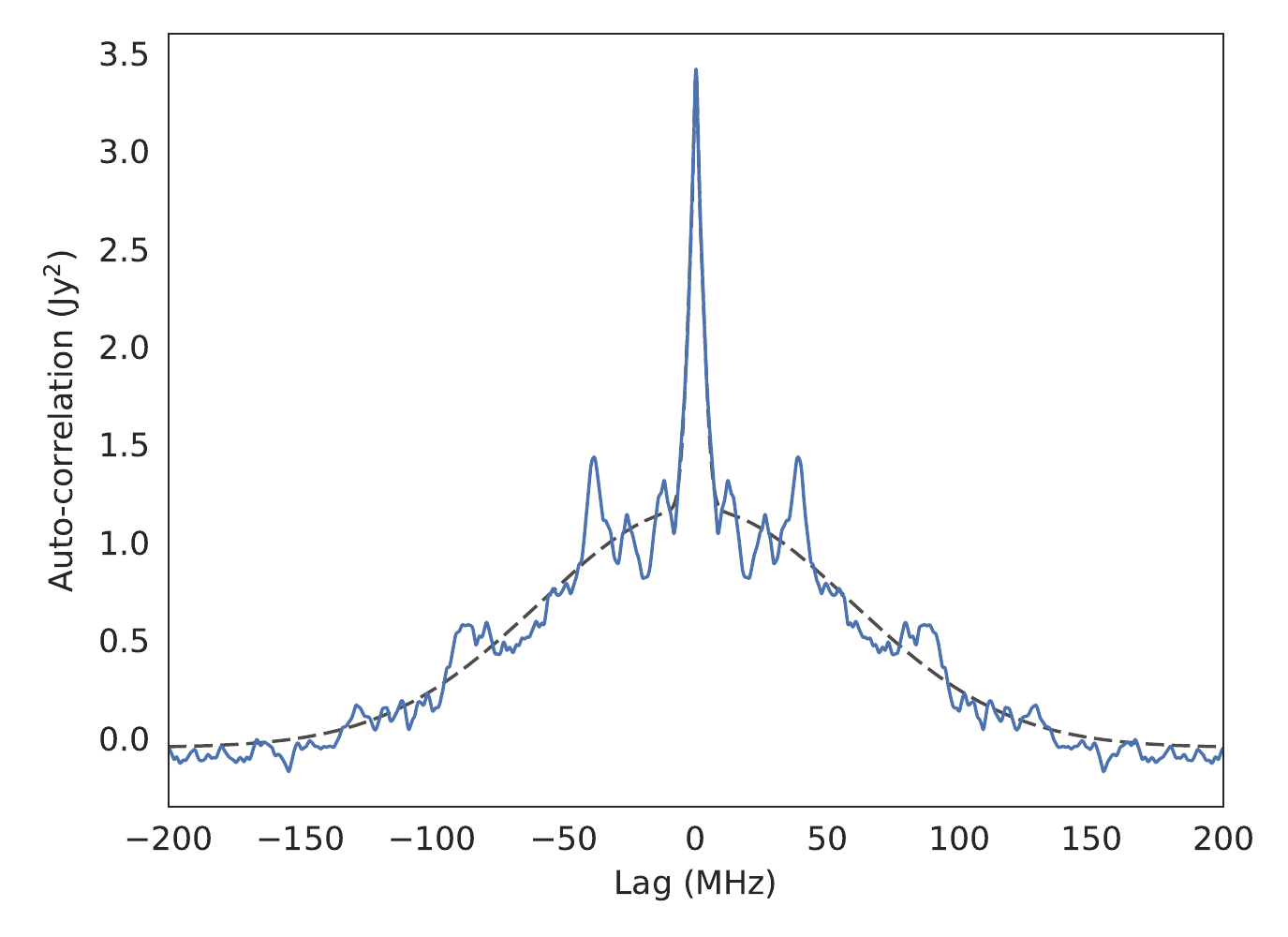}
    \caption{Auto-Correlation function of the time-averaged spectrum. There is a
    minimum frequency lag scale of $\sim6.2$~MHz and maximum of $\sim138$~MHz
    which defines the extent of the spectrum (black dashed).
    }
    \label{fig:acf}
\end{figure}

Figure \ref{fig:acf} shows the \gls{acf} of the time-averaged spectrum over the
extent of the W10 pulse width (Figure \ref{fig:dynamic_spectrum}, right plot). A
two Gaussian component model is fit to the \gls{acf}. The central peak
\gls{fwhm} is $\sim6.2$~MHz which is the characteristic frequency scale of the
bright structures seen in the spectrum.  The band extent of the observed pulse
is $\sim138$~MHz determined by the \gls{fwhm} of the second model component.
The fit scattering timescale $\tau_{\textrm{scatter}}$ indicates a scintillation
bandwidth $\Delta \nu_{\textrm{d}}$ of $\approx 0.25$~kHz
\citep{1998ApJ...507..846C}, a much smaller scale compared to the measured
characteristic size of the spectral structures. The NE2001 line of sight model
\citep{2002astro.ph..7156C} predicts a scattering timescale of
\SI{2.2}{\micro\second} and scintillation bandwidth of 50\,kHz, two orders of
magnitude off from the fit size scales.  The spectrum structure could be
intrinsic to the source or due to a complex intervening medium, but we are
cautious not to over-interpret this result. A single pulse provides insufficient
information to build a scintillation model.

\subsection{Polarization and Rotation Measure}

The frequency-integrated pulse profile (Figure \ref{fig:dynamic_spectrum},
bottom plot) shows little to no polarization structure but the spectrum (Figure
\ref{fig:dynamic_spectrum}, right plot) contains features with significant
linear and circular polarization.

\begin{figure}
    \includegraphics[width=1.0\linewidth]{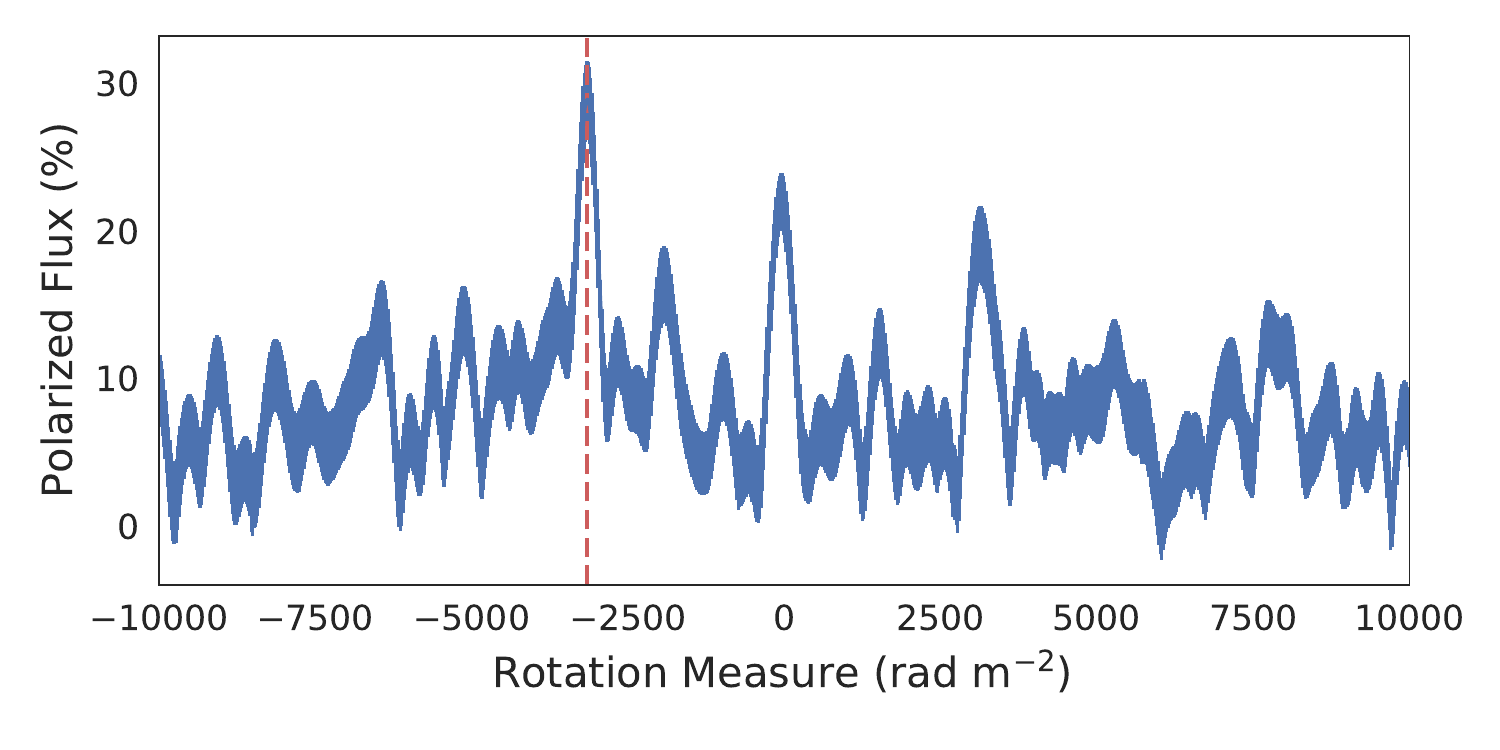}
    \caption{Linear polarization fraction as a function of rotation measure
    computed during a brute-force rotation measure fit. The peak is at
    $-3163$~rad~m$^{-2}$.
    }
    \label{fig:rm_spectrum}
\end{figure}

We attempt to fit a  Faraday rotation model to the spectrum to 
account for the observed linear polarization.  
We used the rotation measure fitting tool
\texttt{rmfit} from \texttt{PSRCHIVE} \citep{2004PASA...21..302H} to find the
peak linear polarization at a rotation measure of $-3163 \pm 20$~rad~m$^{-2}$.
Figure \ref{fig:rm_spectrum} shows the linear polarization flux as a function of
rotation measure. There are significant peaks at $0$~rad~m$^{-2}$ and
$+3163$~rad~m$^{-2}$ indicating Faraday rotation is possibly not a good model to
the observed frequency-dependent polarization response.

\begin{figure*}
    \includegraphics[width=1.0\linewidth]{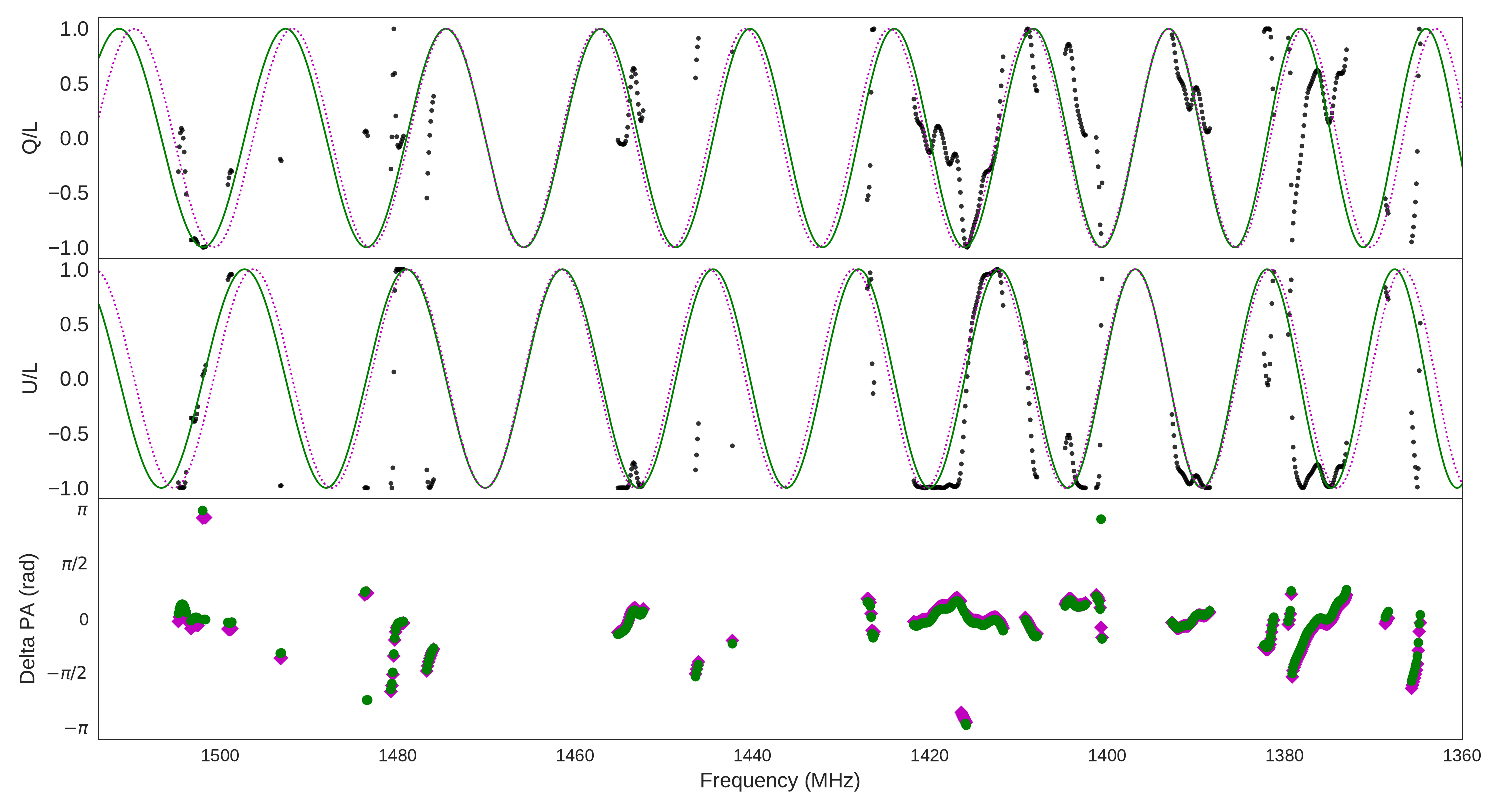}
    \caption{Top and centre plots show best fit from QU-fitting using a
    $\lambda^2$--relation (solid green), and $\lambda$--relation (dashed
    magenta) for channels with an \gls{snr} $> 5$ (black circles). Bottom plot
    shows the residual polarization angle for the $\lambda^2$--relation model
    (green circles) and $\lambda$--relation model (magenta diamonds).
    }
    \label{fig:QUfit}
\end{figure*}

We explored the Faraday rotation model further by performing a similar
QU-fitting analysis as that presented in \cite{2018Natur.553..182M}.  We used a
Faraday rotation model $\textrm{PA}_{\textrm{Faraday}}(\lambda) = \textrm{PA}_0
+ \textrm{RM} \, \lambda^2$, and the normalized Q and U values were fit for
simultaneously. For reference, a lower order model that scales linearly with
wavelength, $\textrm{PA}_{\textrm{Linear}}(\lambda) = \textrm{PA}_0 + \textrm{L}
\, \lambda$, was also fit.  Figure \ref{fig:QUfit} shows the optimal fit for a
Faraday rotated model (green) and a simple $\lambda$--relation (magenta).  Only
regions of the spectrum where the Stokes I exceeds an \gls{snr} of 5 were used
for the fitting.  A brute-force fit of the parameters in both models was
performed, resulting in a Faraday rotation model fit of $\textrm{RM} = -3156,
\textrm{PA}_0 = 0.088$ and a linear modal fit of $\textrm{L} = -1322,
\textrm{PA}_0 = 0.321$.  Both models result in $\chi^2$ residuals that are
within $5\%$ of each other.  The residual polarization angle across the band
(bottom plot) shows that neither model completely accounts for the observed
polarization structure. It is likely the polarization structure is not due to
Faraday rotation alone.

We consider if this frequency-dependent structure is due to a poor polarization
calibration. Polarization calibration was performed by using a noise diode
reference observation, a standard process when observing with Parkes. We found
no frequency-dependent polarization excess. This polarization calibration
assumes an ideal feed model for a source detected at the beam centre. Since the
location of the detection in the beam was unknown this polarization structure
could be due to a frequency-dependent instrumental polarization leakage.
\cite{2011ITAP...59.2058C} note Parkes has low polarization leakage across a
portion of the multibeam band that was measured, but again, this was only
reported for the beam centre. If the frequency-dependent polarization structure
is instrumental this would indicate the source was located far from the beam
centre. The multibeam receiver far side-lobes are not well modelled, as such, we
are uncertain if they would induce such a characteristic frequency-dependent
structure.

\subsection{Cyclostationary analysis}

Modulation schemes employed in modern communication exhibit cyclostationary
features \citep{Gardner:2006}. To search for evidence of signal modulation that
would suggest terrestrial origin---or indeed, emission from a technologically
advanced extraterrestrial civilization---we performed cyclic spectroscopy
\citep{ANTONI2007597,ANTONI2009987} on the coherently de-dispersed pulse. To
maximize S/N, we extracted 10\,ms of data around the FRB event from the
brightest 3.5\,MHz coarse channel (1459.5--1463 MHz). We then computed the
cyclic spectral density for the Y-polarization, in which the signal was
strongest. No cyclic features were apparent.

To verify this approach, we simulated a transient \gls{bpsk} signal at the
apparent \gls{snr} of the \gls{frb}, and repeated the analysis.  No cyclic
features were seen, which suggests that the \gls{snr} of the \gls{frb} is not
high enough to preclude communication emissions exhibiting cyclostationary
features. At higher \gls{snr} (+20 dB, i.e. 100x), 
the cyclic features are indeed apparent.

\section{Follow-up Observations}\label{sec:follow-up}

After post-detection calibration procedures, we observed FRB\,180301 for a
further 82 minutes to search for repeated bursts. The BPSR real-time detection
system did not report any burst candidates during this period. In order to
perform a deeper search, we generated Stokes-I filterbank files from the
\gls{bl} data with time and frequency resolution of \SI{75}{\micro\second}
and 0.435\,MHz.
We then searched for pulses over 1207.5--1361.5\,MHz and also over
1361.5--1515.5\,MHz, with DM range 1--2000\,pc\,cm$^{-3}$ using the
\textsc{Heimdall} package \citep{Barsdell:2012}. We visually inspected dynamic
spectra surrounding each candidate event with a \gls{snr}$ > 6$ (8661
candidates), but found no significant events similar to FRB\,180301.

As reported in \citet{Liping:2018atel}, we also performed optical 
follow-up observations of FRB\,180301 with the 1.35--m
SkyMapper telescope in $g$-- and $r$--band \citep{2007PASA...24....1K} using an
email-based triggering mechanism. SkyMapper has a wide
field-of-view (5.7 square degrees) that fully covers the Parkes localization
region of FRB~180301 ($\sim$14 arcmin beam size). This automatic response
resulted in a sequence of ten 100-second exposures initiated about 3.2 hours
after the burst. The first image was obtained at 10:48:28 UTC on 2018 March 1
and the subsequent nine images were slightly dithered to fill in the gaps
between CCDs.

\begin{figure}
    \includegraphics[width=1.0\linewidth]{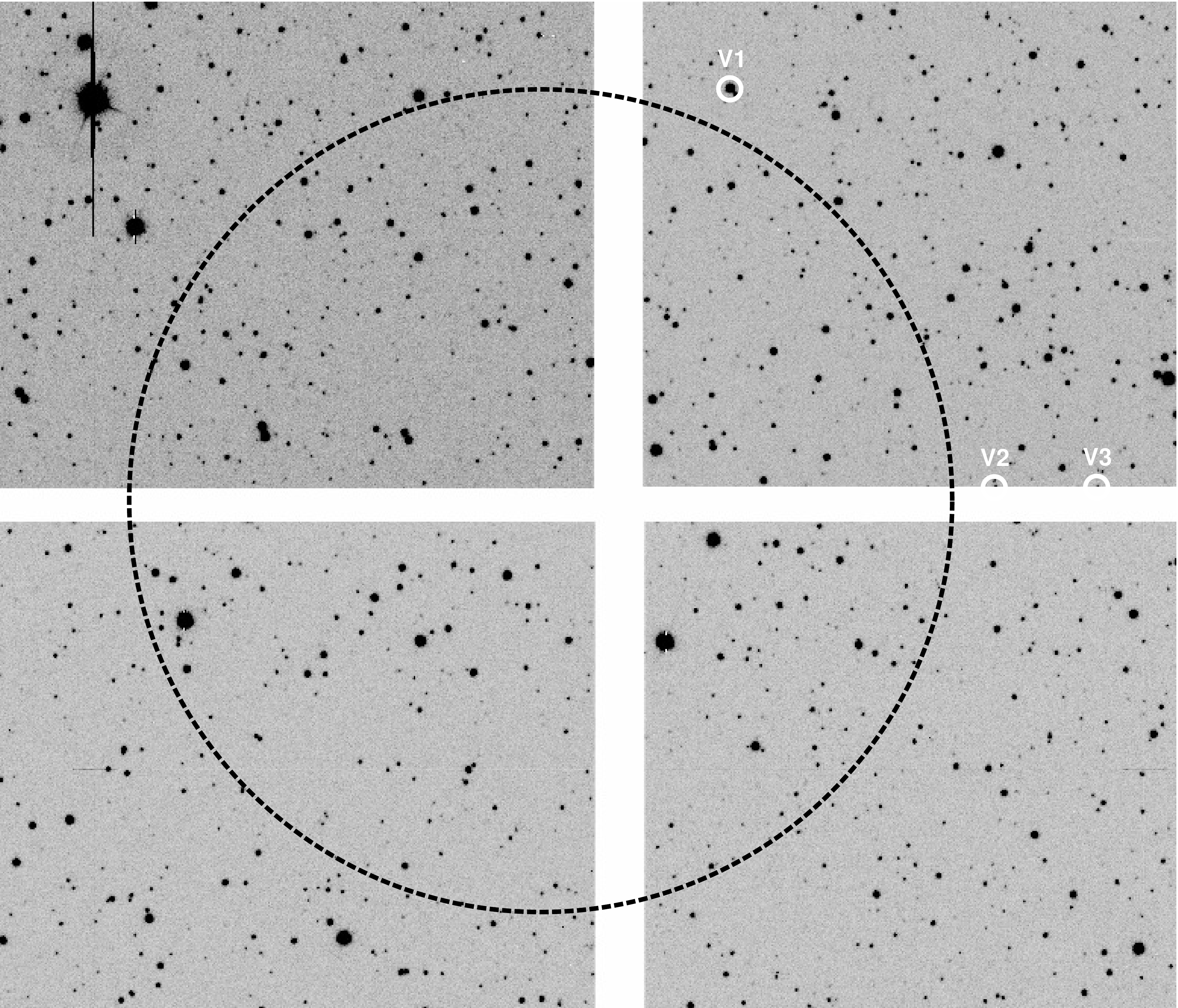}
    \caption{First SkyMapper $r$-band image at the position of FRB~\,180301. The
    black circle represents the beam size of the Parkes radio telescope. White
    circles indicate three stars detected as variable during the observations,
    although these are not associated with the FRB.}
    \label{fig:skymapper}
\end{figure}

We searched for optical transient candidates within the 14\,arcmin Parkes beam
using the SkyMapper transient detection pipeline described in
\citet{2017PASA...34...30S}. A deeper, co-added reference image was taken on
2018 March 9, in order to carry out image subtraction in an optimal manner.
Since there is as yet no available calibration sources from the first data release of
the SkyMapper Southern Survey \citep{2018PASA...35...10W}, we use the APASS
catalogue (AAVSO Photometric All Sky Survey: \citealt{2016yCat.2336....0H}) to
estimate {95\%}--confidence magnitude limits, resulting in point sources
detected down to $r\sim 19.4$ and $g\sim 19.2$ mag, which is limited by sky
brightness from the moon. We find no transient or variable sources within a
7--arcmin radius of the Parkes beam center in any of the resultant images (see
Fig. \ref{fig:skymapper}).

We triggered observations at the position of FRB180301 in the $BVgri$ bands with
the Las Cumbres Observatory \citep{2013PASP..125.1031B} 1--meter telescope
network at 13:22 UTC on 2018 March 1. The first images were obtained at 18:06
UTC on 2018 March 1 with one of the 1--meter telescopes at the South African
Astronomical Observatory. We find no new sources in the images when performing a
visual comparison to archival Digitized Sky Survey images.  The images are
available for download through the LCO
Archive\footnote{\url{http://archive.lco.global}} by searching for object
`FRB180301'.

Several follow-up observations of FRB\,180301 were reported by other facilities.
\citet{Anumarlapudi:2018atel2} report no evidence for any hard X-ray transient
within the energy range of 20--200~keV. \citet{Savchenko:2018atel} report no
significant GRB counterpart, estimating a 3-sigma upper limit on the
75--2000~keV fluence of 4.0$\times10^{-7}$\,erg\,cm$^2$ for a sub-second with a
characteristic short GRB spectrum occurring within 300\,s of the FRB\,180301
detection.

\section{Discussion}\label{sec:discuss}

FRB\,180301 does not fit the prototypical model of an \gls{frb}, as shown in the
verification heat map (Figure\,\ref{fig:heatmap}). Aside from concerns on the
RFI environment, the polarization structure is unusually complex, the pulse
appears band limited, we find no preferred pulse profile model, and given the
complex spectral structure a spectral index model cannot be fit.  Given the
ambiguity and atypical features exhibited by FRB\,180301, we discuss some
concerns, and potential anthropogenic mechanisms, below.

\subsection{Anthropogenic or Astrophysical?}

The presence of \gls{gps} L3 emission during the period of the burst, along with
the complex frequency and polarization structure of the pulse, give ambiguous
evidence for either an astrophysical or anthropogenic origin.

Also of concern is that three other FRB events, FRB\,180309
\citep{Oslowski:2018frb1}, FRB\,180311 \citep{Oslowski:2018frb2}, and
FRB\,180318 (Oslowski, p.c.) were detected with Parkes within 17 days of
FRB\,180301, which is statistically anomalous. Based on an event rate of
$1.7\pm^{1.5}_{-0.9}\times10^{3}$ events sky$^{-1}$ day$^{-1}$  above a
2\,Jy\,ms fluence \citep{Bhandari:2018}, we calculate an expectation value of
0.06--0.25 events over the $\sim$140 hours of FRB observations during MJD
58178--58196. The corresponding Poissonian probability $P(N\geq4)$ is
$1.3\times10^{-4}$.  Nevertheless, while the bunching of events is improbable,
one cannot conclusively state that one or more of these bursts is spurious. 

An anthropogenic pulse origin does potentially solve some
open questions. The band-limited pulse is consistent with an antenna
transmission model. The complex polarization structure, which does not appear to
follow a Faraday model, could be due to signal modulation. The non-repeating
nature is explained as a satellite
would be moving and these transmissions are rare. The scattering and scintillation
time-scale discrepancy is thus explained as the signal is neither scattered nor
scintillating.

Nevertheless, there are several arguments against FRB\,180301 being
\gls{gps}-related \gls{rfi}. Firstly, the \gls{gps} transmission bands are well
defined, and do not extend over the band of the detected pulse (although it could 
be a low-power emission from a malfunctioning or unreported sub-system). Unlike radar
systems, \gls{gps} satellites are not known to emit chirped pulses that could be
mistaken for dispersion.  Secondly, the L3 emission was detected in all beams,
while FRB\,180301 was only detected in a single beam.  GPS satellites are
ubiquitous, so one might further expect such a signal to have been detected
previously, or previous detections have been erroneously reported as \glspl{frb}. 
Finally, the pulse is dispersed, which could be due to a radar system, but no 
such system is present on \gls{gps} satellites.

One possible explanation to FRB\,180301's origin is that it is a ground-based 
L-band (1--2\,GHz) wavelength radar reflection used for range-finding during 
the \gls{gps} L3 emission testing.  \gls{arsr} \citep{4241738,2012ITAES..48..103W}, 
military radar and telemetry is known to operate at L-band frequencies. 
\gls{gps}  was originally a \gls{usaf} technology developed for military use. 
A wide-band pulse
is common in long-distance range finding; as the observed pulse is band-limited,
its duration is consistent with a maximal-power transfer from a ground-based
mono-static radar to an object in mid-Earth orbit. Wide-band, dispersed pulses
are known to exist and have been previously detected in \gls{frb} search
pipelines \citep{2018MNRAS.481.2612F}.  The reflected pulse would appear much
weaker than the L3 emission, but still dispersed. 

Any anthropogenic explanation also needs to explain the frequency-dependent 
polarization evident in FRB\,180301. Polarimetric imaging radar uses 
polarized pulses to measure the characteristic of surfaces based on scattering,
and L-band polarimetric imaging radar systems are known to exist 
\citep[e.g.][]{Gray:2011}. 

\citet{Katz:2016} argues that space-based radar is unlikely to be the origin of
\glspl{frb} since the broad range of dispersion measures and pulse characteristics
would suggest an implausible number of space-based radar systems (or systems
with a peculiar variety of chirp rates),  but does not discuss reflections from
ground-based radars. We note that the variety in pulse characteristics could be
due to deliberate signal obfuscation. Also, the existence of a population of FRB 
events with broadly ranging characteristics does not mean that one or more events are
spurious, as is the case of the simulated pulses and the events detected by the
25-m Nanshan Telescope  reported in \citep{2018MNRAS.481.2612F}.  

\citet{Kulkarni:2014} note that the chance reflection of a solar flare off a
satellite or the Moon could potentially produce an FRB-like event.  Using the
solar activity database of \citet{Sadykov:2017}, we searched for coincident
solar flare events; no flares were found, and as such we discount this
possibility.

\subsection{Instrumental or intrinsic polarization?}

Faraday rotation can not account for the observed circular polarization in the
spectrum. Assuming the pulse is astrophysical,
the frequency-dependent circular polarization structure indicates
either there is significant instrumental polarization leakage, or that the source
is intrinsically polarized. 

If the source was detected far from the beam centre, it is possible that
instrumental polarization leakage is introducing the frequency-dependent
structure.  However, measurements of the beam response show the first side-lobe
level to be below -25\,dB \citep{2018MNRAS.474.1900M}, and as such the intrinsic
luminosity of the source would be over two orders of magnitude higher than if
located the beam centre, closer to the luminosities for ASKAP FRBs as reported
in \citet{Shannon:2018}. However, no detection is made in inter-beam
correlations, meaning that a particularly advantageous (and unlikely) side-lobe
response in beam 03 is required.  If instrumental polarization leakage is
introducing a frequency-dependent structure, then a rotation measure fit is not
possible without knowledge of where in the beam the event occurred, and an
accurate model for the beam response at that point.

If the source occurred in the primary lobe, then the frequency-dependent 
polarization structure is likely intrinsic to the source. In this case, the 
polarization structure may provide information about the underlying 
emission mechanism of the source. However, as no other FRB reported to date 
has exhibited similar polarization structure, we caution against over interpretation.  
Whatever the origin, an observer should be careful when applying a Faraday rotation 
model when the location of the detection in the beam is unknown.

\subsection{Comparisons with other FRBs}

If FRB\,180301 is astrophysical and the polarization structure is primarily due to Faraday rotation, then it bares similarity to FRB\,121102 with
its complex spectrum and large \gls{rm}. Though, the fit \gls{rm} is
significantly smaller than that of FRB\,121102, no other \gls{frb} has a
similarly large \gls{rm}. The magnetic field strength can be estimated to be
$\langle B_{\mathbin{\|}} \rangle = -8 \mu G$ along the line of sight using
Eq. 4 of \cite{2006ApJ...642..868H}.  This is larger than the mean measured
large-scale Galactic field strength, but consistent with a high \gls{dm} source,
indicating it is not embedded in a similar environment to FRB\,121102
\citep{2018Natur.553..182M}.

FRB\,180301 has similar spectral characteristics to FRB\,170827
\citep{2018MNRAS.478.1209F} in that a single, narrow-band component is the
dominant contributor to the flux, with lower flux structure spread over a
portion of the band. Additionally, there is similar fine structure in the
spectrum. Nonetheless, the FRB\,180301 profile is wider than FRB\,170827, and
the events are detected at different frequencies. As only a single polarization 
was recorded for FRB\,170827, its polarization properties are unknown. 

FRB\,180301, along with FRB\,170827 and FRB\,121102, are the only events for
which complex-voltage data have been captured and reported upon thus far. 
As other detections used
incoherent dedispersion, and in many cases were discovered in 2-bit data products, it
could be the case that most \glspl{frb} do indeed have a complex spectrum
intrinsic to the source and/or due to scintillation, which is only now becoming
apparent with our ability to capture voltage data and perform coherent
de-dispersion. Alternatively, it could be that there are multiple observational
classes of \glspl{frb}: one class that fits the prototypical model without
complex frequency structure \citep{2018MNRAS.481.2612F}, and another class that
exhibits complex structure. Repeating sources may also turn out to be a distinct
class of event.

\section{Conclusion}\label{sec:conclude}

We have reported on the detection of FRB\,180301, a highly-polarized FRB that
exhibits complex frequency structure. Other than FRB\,121102, the detection of
FRB\,180301 is the most complete in terms of addition information for an \gls{frb}
captured to date.
This has allowed a detailed analysis of the coherently dedispersed pulse and its
polarization characteristics.

We performed a rigorous set of tests to verify FRB\,180301 as
astrophysical, but we are unable to
definitively state that FRB\,180301 is not related to human activity. 
Applying
coherent de-dispersion to the signal has revealed complex structure, but it is
unclear that this structure should be attributed to astrophysical origin. Of
particular concern in this instance is the proximity of the event to the
geosynchronous orbit belt, and that a GPS testing campaign is known to have been
conducted during the day.  While circumstantial, the statistically unlikely
detection of four \glspl{frb} at Parkes within a 17-day period is also
troublesome. 

While we approach our classification of FRB\,180301 with caution that it may be
\gls{rfi}, we do not suggest that all \glspl{frb} are anthropogenic.
Observations of FRB\,121102 have shown, unequivocally, that its emission is
astrophysical.  The existence of a population of FRBs, detected at multiple
telescopes, all displaying astrophysical characteristics, remains strong
evidence that FRBs are genuinely an astrophysical phenomena. Nevertheless, as
our analysis shows, without precise localization capabilities, conclusive
verification of a single event remains challenging. 

Interferometric and
multi-site detection are essential to rule out satellites as FRB progenitors.
Arrays with Fresnel zones farther out than \gls{geo} orbits (36000\,km)---which
at 1.4\,GHz corresponds to 2\,km baselines---will be capable of precluding
satellites, or chance reflections off space debris, as sources of FRB-like
\gls{rfi}. ASKAP, MeerKAT, and other upcoming instruments meet this criterion.

The complex voltage data and intermediate data products are publicly hosted at
the Breakthrough Listen data center. Jupyter notebooks with our analysis are
hosted on our public git
repository\footnote{\url{https://github.com/griffinfoster/frb180301-analysis}}.

\section*{Acknowledgements}

We thank Jim Cordes, Jason Hessels, Simon Johnston, Dan Mertely, and Allison
Rung for their comments and help.  Breakthrough Listen is managed by the
Breakthrough Initiatives, sponsored by the Breakthrough Prize Foundation. The
Parkes radio telescope is part of the Australia Telescope National Facility
which is funded by the Australian Government for operation as a National
Facility managed by CSIRO.  SkyMapper is owned and operated by The Australian
National University's Research School of Astronomy and Astrophysics; national
facility capability is funded through Australian Research Council LIEF grant
LE130100104.  Parts of this project were conducted by the Australian Research
Council Centre of Excellence for All-sky Astrophysics (CAASTRO), through project
number CE110001020.  Support for IA was provided by NASA through the Einstein
Fellowship Program, grant PF6-170148.



\bibliographystyle{mnras}
\bibliography{references}


\begin{appendices}

\section{Breakthrough Listen}\label{BL}

\gls{bl} is a ten-year initiative directed at detecting technosignatures that
would indicate the presence of advanced life beyond Earth \citep{Worden:2017aa}.
The initial \gls{bl} program uses the 100-m Robert C.  Byrd Green Bank telescope
in West Virginia, USA, and the 64-m CSIRO Parkes radio telescope to observe a
selection of 1709 nearby stars and 100 nearby galaxies, along with surveying the
Galactic plane \citep{2017PASP..129e4501I, 2017ApJ...849..104E}. In addition,
the 2.4-m Automated Planet Finder optical telescope is also being used, to
conduct a search for narrow-band optical transmissions from targets within the
1709-star sample. Combined, the observations from these telescopes constitutes
the most comprehensive search for technosignatures to date.

In the initial years of the program, 25\% of the total observing time of the
Parkes telescope is assigned for \gls{bl} activities. Observations are typically
scheduled for 10-11 hours per day, up to 4-5 times per week.
A major component of the
\gls{bl} program at Parkes is a 21-cm wavelength Galactic plane survey, which
utilizes the Parkes multibeam receiver \citep{StaveleySmith:1996mb}. The survey
covers Galactic latitudes $|b|<6.5^{\circ}$ over the range of Galactic
longitudes accessible with Parkes, $-174^{\circ}<l<60^{\circ}$. Similar to the
\gls{htru} survey \citep{Keith:2010htru}, a step-and-stare approach with
5-minute pointings is employed. Each beam of the multibeam receiver is separated
by 14\,arcmin in one plane, and $\sqrt{3}\times 14$\,arcmin in the other plane.
As the \gls{fwhm} beamwidth of the receiver is $\sim$14\,arcmin at 21-cm
wavelength, interleaved pointings allow for the survey area to be efficiently
covered with tessellated pointings (see Figure\,2 of \citealt{Keith:2010htru}).

While the \gls{bl} survey follows a similar observational strategy to the
\gls{htru} and \gls{superb} surveys \citep{Keane2015}, technosignature searches
require a far higher spectral resolution ($\sim1$\,Hz,
\citealt{2015aska.confE.116S})  than that available in archival data products
from the HTRU/SUPERB surveys ($\sim390$\,kHz).  As such, new digital recorder
systems have been installed at both the Parkes and Green Bank observatories to
allow voltage capture to disk across the full bandwidth of the available
receivers \citep{2018PASP..130d4502M,2018PASA...35...41P}.

\end{appendices}

\bsp	
\label{lastpage}
\end{document}